\documentclass[twocolumn]{book}
\usepackage[dvips]{graphicx,color}
\usepackage{makeidx,universe}

\makeauthorindex

\BookTitle{Proceedings of the XXIX PHYSICS IN COLLISION}

\CopyRight{\copyright 2009 by Universal Academy Press, Inc.}

\begin{document} 

\pagenumbering{arabic}

\chapter{%
{\LARGE \sf
Probing QCD at the Highest $Q^2$ Deep Inelastic Scattering} \\
{\normalsize \bf 
James Ferrando} \\ 
{\small \it \vspace{-.5\baselineskip}
  University of Oxford, 
    Particle Physics,
	Denys Wilkinson Building,
Keble Road, Oxford, United Kingdom
OX1 3RH.
       \\
}
}

\AuthorContents{J.\ Ferrando}

\AuthorIndex{Ferrando}{J.}

  \baselineskip=10pt 
  \parindent=10pt    

\section*{Abstract} 

Recent results from the HERA $ep$ collider are reviewed in these proceedings. The results
are from measurements that probe QCD at high-energy scales, as defined by $Q^2$, the four-momentum-transfer squared of the collisions. These cross-section measurements provide information about the parton
distribution functions (PDFs) of the proton and can be used to constrain global
fits of these PDFs. Recent measurements of the strong coupling $\alpha_S$ and jet substructure from HERA at similar energy scales are also reviewed.

\section{Introduction} 

The Hadron-Elektron Ringanlage (HERA) was the first and, so far, only electron-proton collider in the world. At HERA, protons were collided with electrons ($e^+$ or $e^-$) at 
the interaction region of the H1 and ZEUS detectors. The collider was operational from
1992 until 2007. The proton beam energy was $820\,\mathrm{GeV}$ from 1994--1997 and then  $920\,\mathrm{GeV}$ until 2007. In the final months of HERA running, proton
energies of $460\,\mathrm{GeV}$ and $575\,\mathrm{GeV}$ were used for the purposes of measuring the longitudinal structure function $F_L$.  The electron beam energy was $27.5\,\mathrm{GeV}$ throughout the whole of HERA running.

Between 2000 and 2003, HERA underwent a luminosity upgrade and was also modified
to enable production of longitudinally-polarised electron beams. The data taken
before (after) this upgrade are referred to as HERA-I (HERA-II) data.

HERA was the location {\it par excellence} for the measurement of proton structure. The HERA experiments extended the kinematic reach of Deep Inelastic Scattering (DIS) experiments by several orders of magnitude in photon virtuality, $Q^2$, and Bjorken-$x$. In addition, the use of a proton beam rather than a nuclear target meant that no nuclear corrections were needed for use in extracting the
parton distribution functions (PDFs) of the proton. The use of both $e^+$ and 
$e^-$ beams means that electroweak effects can be exploited for sensitivity to
cross-section contributions from different quarks.

The electron-proton collision environment is also ideal for jet physics. Jets
were copiously produced at HERA and underlying event activity is much less
significant in electron-proton collisions than in hadron-hadron collisions. Jet production cross sections are sensitive
to the strong coupling, $\alpha_S$, and the gluon PDF. The large quantity of 
jets produced means that the substructure of jets can also be studied.

\section{Structure Functions and PDFs}

The double-differential cross section for neutral current (NC) DIS with unpolarised $e^{\pm}p$ beams
can be written~\cite{ferrando:dcs} in terms of $Q^2$, $x$ and the electron inelasticity, $y$, as:

\begin{eqnarray}
\frac{\mathrm{d}^2\sigma^{e^{\pm}p}_{NC}}{\mathrm{d}x \mathrm{d}Q^2} & = & \frac{2 \pi \alpha^2  Y_{+}}{xQ^4}[F_2(x,Q^2) \\ \nonumber
&  & \mp \frac{Y_{-}}{Y_{+}}xF_3(x,Q^2)\\
 &  & - \frac{y^2}{Y_{+}}F_{L}(x,Q^2) ]\ (1+\delta_r) \nonumber
\end{eqnarray}

where $F_2$, $xF_3$ and $F_L$ are the proton structure functions, $\delta_r$ is an electroweak radiative correction, and $Y_{\pm} = 1\pm (1-y)^2$.  

At leading order in Quantum Chromodynamics (QCD), $F_2 \propto  \sum_i \{ xq_i(x,Q^2) + x\bar{q}_i(x,Q^2)\}$ where $q_i$ ($\bar{q}_i$) are the parton distribution functions for the (anti)quarks. This $F_2$ term dominates the cross section over the majority of the kinematic region. The parity violating structure function $xF_3$ is proportional to $ \sum_i \{ xq_i(x,Q^2) - x\bar{q}_i(x,Q^2)\}$, i.e. the valence quarks. This $xF_3$ term is significant only at high values of $Q^2$. At leading order in QCD the longitudinal structure function $F_L$ is zero. However, once higher order terms are taken into account, it can be seen that $F_L$ has a value proportional to the gluon PDF and the strong coupling. The $F_L$ term becomes significant at low values of $Q^2$ and high values of $y$.

\subsection{Measurements of $F_2$ and $xF_3$}
\label{ferrando:sec:nc}

\begin{figure}[t]

   \begin{center}
     \includegraphics[width=.50\textwidth]{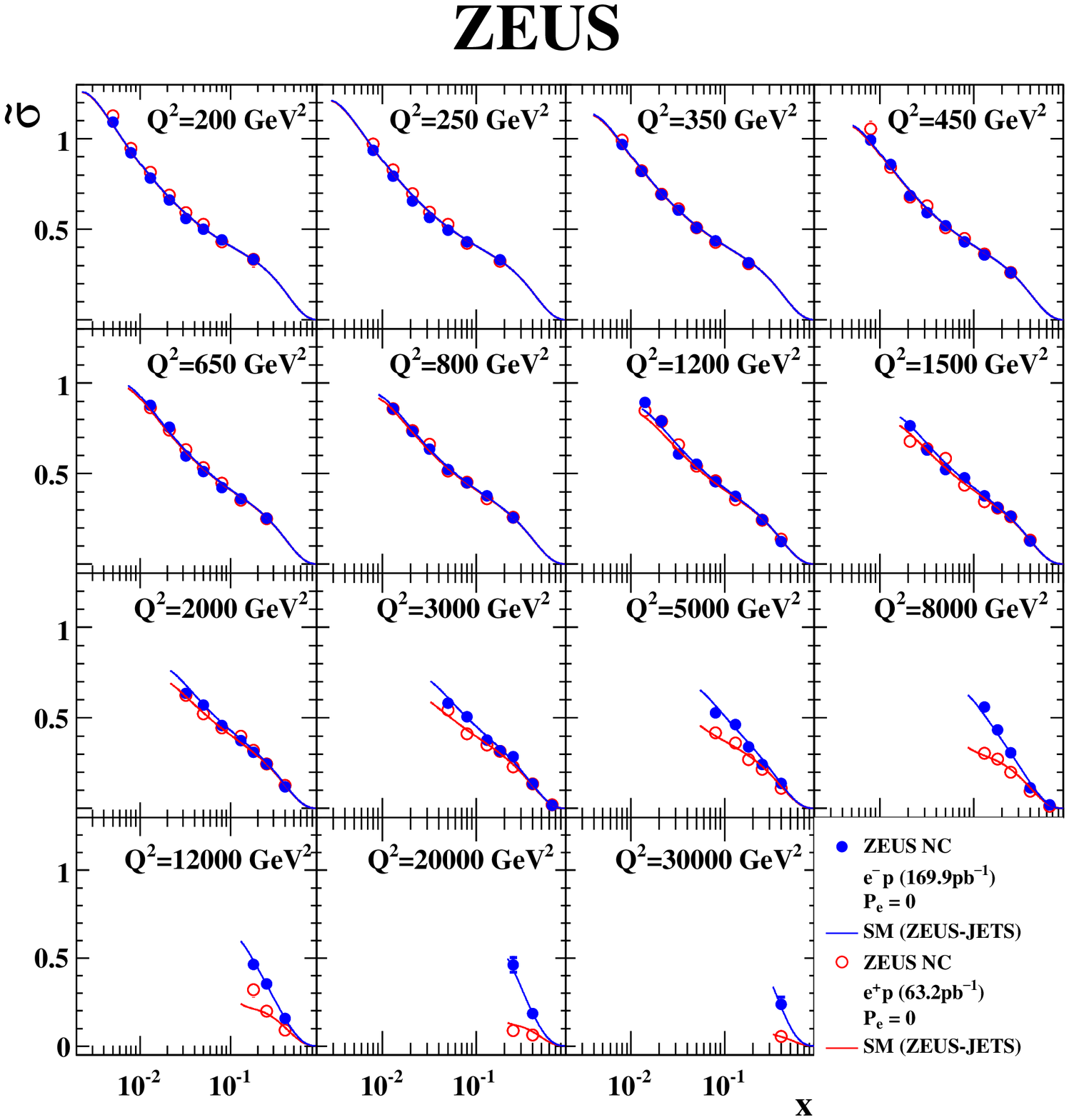}
      \caption{The reduced cross sections for NC DIS in $e^{\pm}$ collisions. }
    \label{ferrando:fig:redsigma}
   \end{center}

\end{figure}

If one divides the double-differential cross section for NC by the ``propagator'' term $\frac{2 \pi \alpha^2  Y_{+}}{xQ^4}$ and the electroweak radiative correction then one obtains $\tilde{\sigma}$, the reduced cross section at Born level:

\begin{eqnarray}
\tilde{\sigma}^{e^{\pm}p}_{NC}  & = & F_2(x,Q^2)    \mp \frac{Y_{-}}{Y_{+}}xF_3(x,Q^2) \label{ferrando:eq:myredsigma} \\
&  & - \frac{y^2}{Y_{+}}F_{L}(x,Q^2) \nonumber
\end{eqnarray}

\begin{figure}[t]

   \begin{center}
     \includegraphics[width=.47\textwidth]{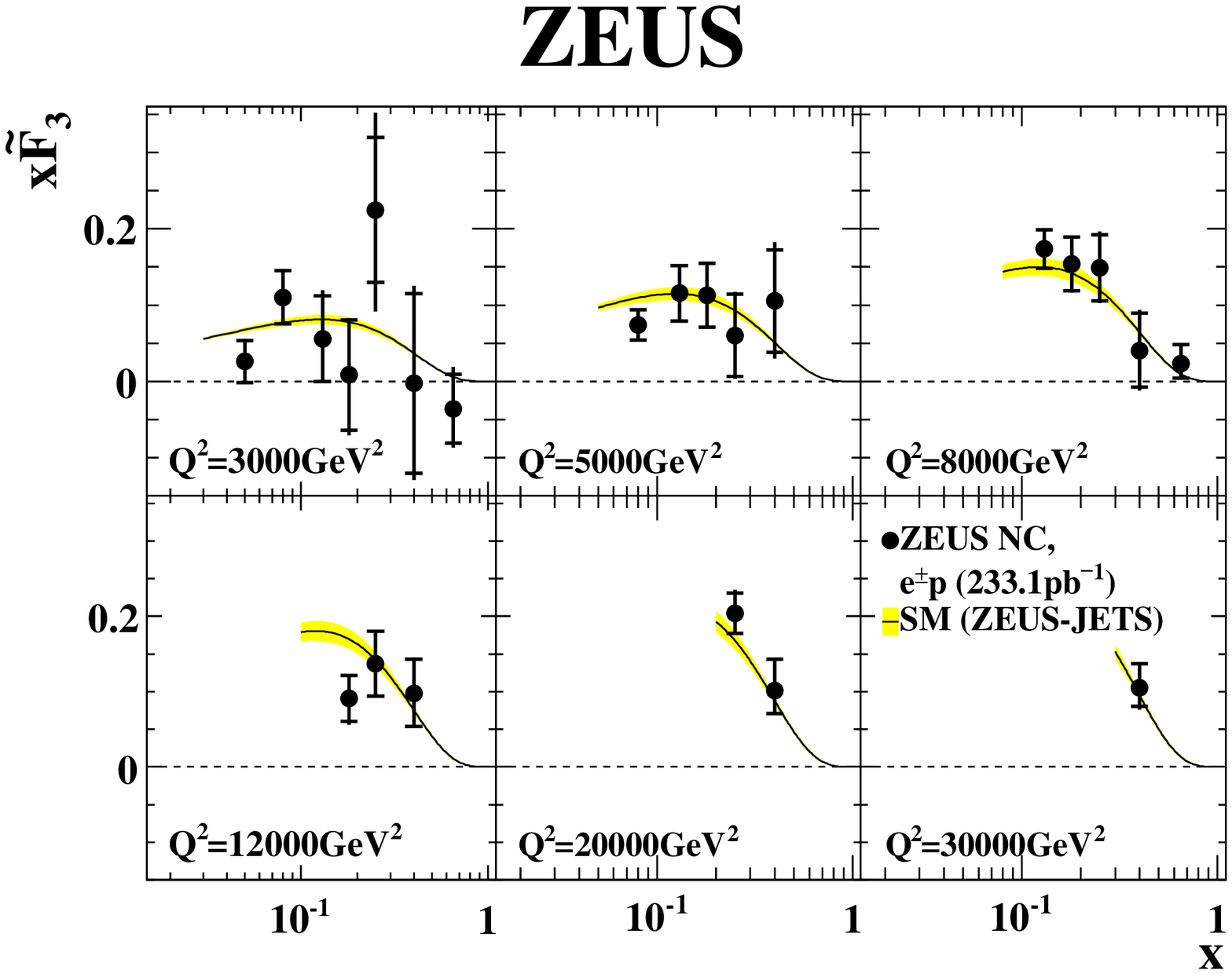}
      \caption{The parity violating structure function $xF_3$ as measured by the ZEUS collaboration. }
    \label{ferrando:fig:xf3}
   \end{center}

\end{figure}

It can be seen that by measuring the reduced cross sections for $e^+p$ and $e^-p$ collisions and subtracting one from the other, it is possible to directly measure $xF_3$. The reduced cross sections for HERA-II $e^-p$ data from a recent measurement by ZEUS~\cite{ferrando:zeushq2nc} are shown in Fig.~\ref{ferrando:fig:redsigma} together with 
data from a previous measurement of the reduced cross sections for $e^+p$ 
data~\cite{ferrando:zeushq2ncold}. The
predictions from the ZEUS-JETS next-to-leading-order(NLO)-QCD fit~\cite{ferrando:zeusjets}, which does not include the $e^-p$ data, are also shown. It can be seen that $\tilde{\sigma}^{e^-p}_{NC}$ and $\tilde{\sigma}^{e^+p}_{NC}$ differ by more at larger values of $Q^2$. This
corresponds to the SM expectation and occurs because the $xF_3$ contribution
(dominated by  interference between $Z$ and $\gamma$ exchange) becomes more significant at higher-energy scales. The reduced cross sections in  Fig. \ref{ferrando:fig:redsigma} were used to extract the values of $xF_3$ shown in  Fig. \ref{ferrando:fig:xf3} This is the most precise direct measurement of $xF_3$ and agrees well with the 
predictions from the ZEUS-JETS QCD fit.

In addition to performing new measurements with HERA-II data, H1 and ZEUS have made considerable progress in the combination of results from HERA-I data~\cite{ferrando:h1zeus:nccc}. In order to produce the combined results, the following strategy is adopted:

\begin{itemize}
\item all measurements are ``swum'' to a common $x$, $Q^2$ grid;
\item the data are corrected to correspond to measurements at the same proton energy ($920\,\mathrm{GeV}$);
\item the average cross sections and uncertainties for the points are calculated with a global combination, described in a recent H1 publication~\cite{ferrando:h1:lowq2};
\item the procedural uncertainties are calculated.
\end{itemize}

\begin{figure}[t]

   \begin{center}
     \includegraphics[width=.47\textwidth]{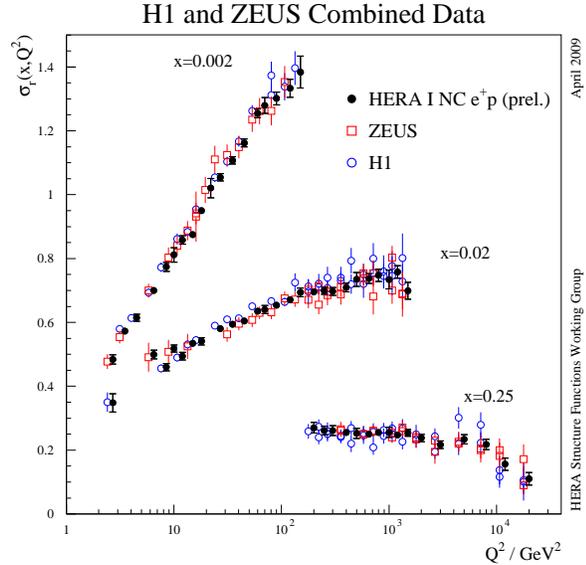}
      \caption{The combined H1+ZEUS cross sections for NC DIS compared to measurements from the individual collaborations. }
    \label{ferrando:fig:h1zeusnc}
   \end{center}
\end{figure}

In total 1402 original points are combined into 741 points. The agreement of the H1 and ZEUS datasets can be quantified by the value of $\chi^2/\mathrm{n.d.f.}$ for the combination. This is $637/656$ indicating good agreement. There are 110 systematic uncertainties considered from the two experiments, and only 3 arising from the combination procedure. An example of the combined data compared to the individual measurements is shown in 
Fig. \ref{ferrando:fig:h1zeusnc}, it can be seen that the new combined measurements are much more precise than the measurements from the individual collaborations. The
combined data attain a precision of $2\%$ or better over a $Q^2$ range of $3< Q^2< 500\,\mathrm{GeV}^2$. Even more impressive precision of $1\%$ or better is attained for $20 <Q^2  <100  \,\mathrm{GeV}^2$.

\subsection{Measurements of $F_L$}

\begin{figure}[t]

   \begin{center}
     \includegraphics[width=.49\textwidth]{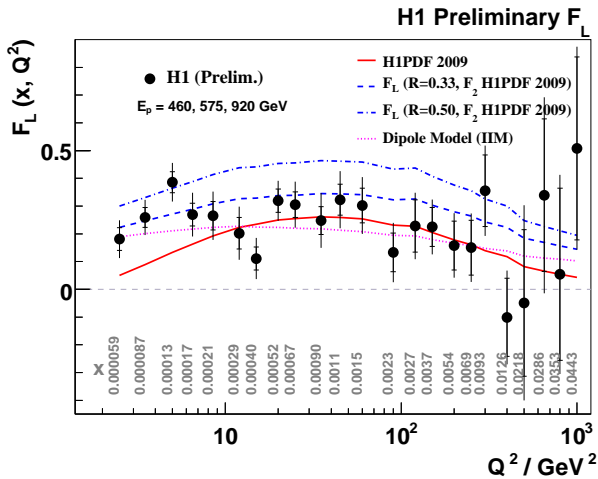}
      \caption{The longitudinal structure function $F_L$ measured by H1.}
    \label{ferrando:fig:fl}
   \end{center}

\end{figure}

From inspection of the expression for the reduced cross section in equation (2)
it can be seen that only the $xF_3$ and $F_L$ terms depend on all of $Q^2$, $x$ and $y$ whereas $F_2$ depends only on $Q^2$ and $x$. It is therefore possible to directly extract $F_L$ by measuring the reduced cross section at different values of $y$ for the same $x$ and $Q^2$. This necessitates a change of centre of mass energy, $\sqrt{s}$ because, at any given $s$, $Q^2$ and $x$ determine $y$ via the relation $Q^2=sxy$. To enable the first direct measurements of $F_L$, HERA dedicated several months of running time at lower proton energies of   $460\,\mathrm{GeV}$ and $575\,\mathrm{GeV}$. The first measurements from H1~\cite{ferrando:h1:fl} and ZEUS~\cite{ferrando:zeus:fl} have now been performed covering the kinematic region $ 12 < Q^2< 90  \,\mathrm{GeV}^2$ and  $ 20 < Q^2< 130  \,\mathrm{GeV}^2$  respectively. The preliminary results of an attempt by H1 to extend the kinematic region to  $ 5 < Q^2< 800  \,\mathrm{GeV}^2$ is shown in Fig. \ref{ferrando:fig:fl} The higher $Q^2$ measurements have much larger uncertainties, dominated by the statistical uncertainty. At
lower $Q^2$ the precision is sufficient to distinguish between theoretical expectations for $F_L$.

\subsection{Charged current cross sections}

\label{ferrando:sec:cc}

\begin{figure}[t]
   \begin{center}
     \includegraphics[width=.49\textwidth]{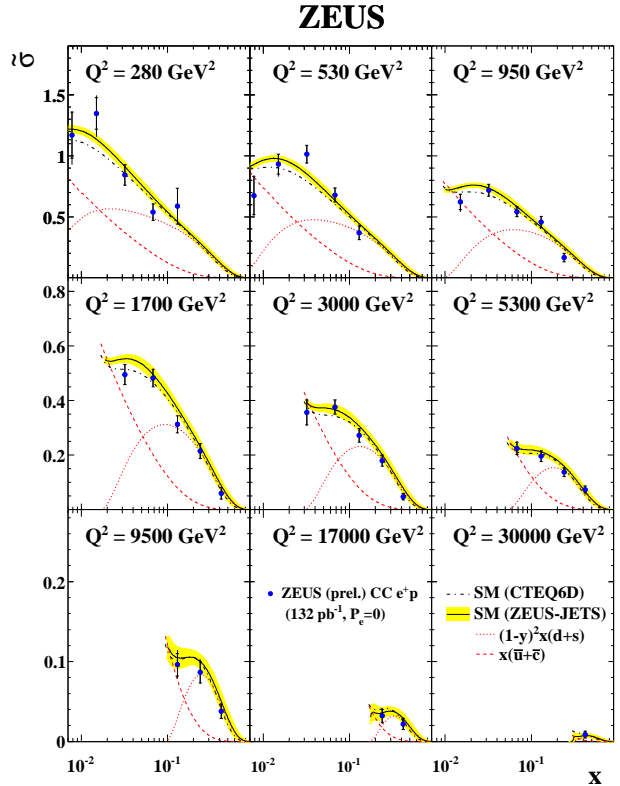}
      \caption{The reduced cross section for CC DIS in $e^+p$ scattering measured by ZEUS.}
    \label{ferrando:fig:cc}
   \end{center}
\end{figure}

The double differential cross section for charged current (CC) DIS can be written as~\cite{ferrando:dcs}:
\begin{eqnarray}
\frac{\mathrm{d}^2\sigma^{e^{\pm}p}_{CC}}{\mathrm{d}x \mathrm{d}Q^2} & = & \frac{ G^2_F M_W^2}{4\pi x(Q^2 +M_W^2)^2}[F_2^{CC}(x,Q^2) \\ \nonumber
 &  & - \frac{y^2}{Y_{+}}F_{L}^{CC}(x,Q^2) \\ 
&  & \mp \frac{Y_{-}}{Y_{+}}xF_3^{CC}(x,Q^2)]\ (1+\delta_r) \nonumber,
\end{eqnarray}

where $M_W$ is the mass of the $W$ boson and the structure functions $F_i^{CC}$ are the structure functions for CC scattering. As with the NC cross section, the propagator $\frac{ G^2_F M_W^2}{4\pi x(Q^2 +M_W^2)^2}$ and radiative correction can be divided out to give the reduced cross section $\tilde{\sigma}^{CC}_{e{^{\pm}}}$. For CC the reduced cross section is proportional (at leading order in QCD) to quark PDFs:

\begin{eqnarray}
\tilde{\sigma}_{e^+}^{CC} & \propto & x[(1-y)^2({d}+s)+\bar{u}+\bar{c}] \nonumber \\ 
\tilde{\sigma}_{e^-}^{CC} & \propto & x[(1-y)^2(\bar{d}+\bar{s})+{u}+c]. \nonumber
\end{eqnarray}

Here, $u$, $d$, $s$, $c$ are the relevant quark PDFs (functions of $x$ and $Q^2$).
This means that measurements of $e^+p$  ($e^-p$) scattering cross sections provide information about the $d$-($u$-)quark PDF. Thus CC cross sections provide
complementary information about the PDFs to NC cross sections; the NC cross sections are dominated by the $u$-quark contribution for both beam charges. 

Data from the latest preliminary measurement of $\tilde{\sigma}_{e^+}^{CC}$
from the ZEUS collaboration~\cite{ferrando:zeus:ccprel} are shown in Fig. \ref{ferrando:fig:cc} They are compared
with predictions from the ZEUS-JETS QCD fit, which does not include these data.
It can be seen, particularly at high values of $Q^2$ and $x$, that the uncertainties on the new ZEUS measurement are comparable to the uncertainty on the ZEUS-JETS prediction. Hence this data will be useful in constraining the $d$ and $s$  PDFs at higher $Q^2$ and $x$.

\subsection{QCD Fits}

\begin{figure}[bt]

  \begin{center}
     \includegraphics[width=.47\textwidth]{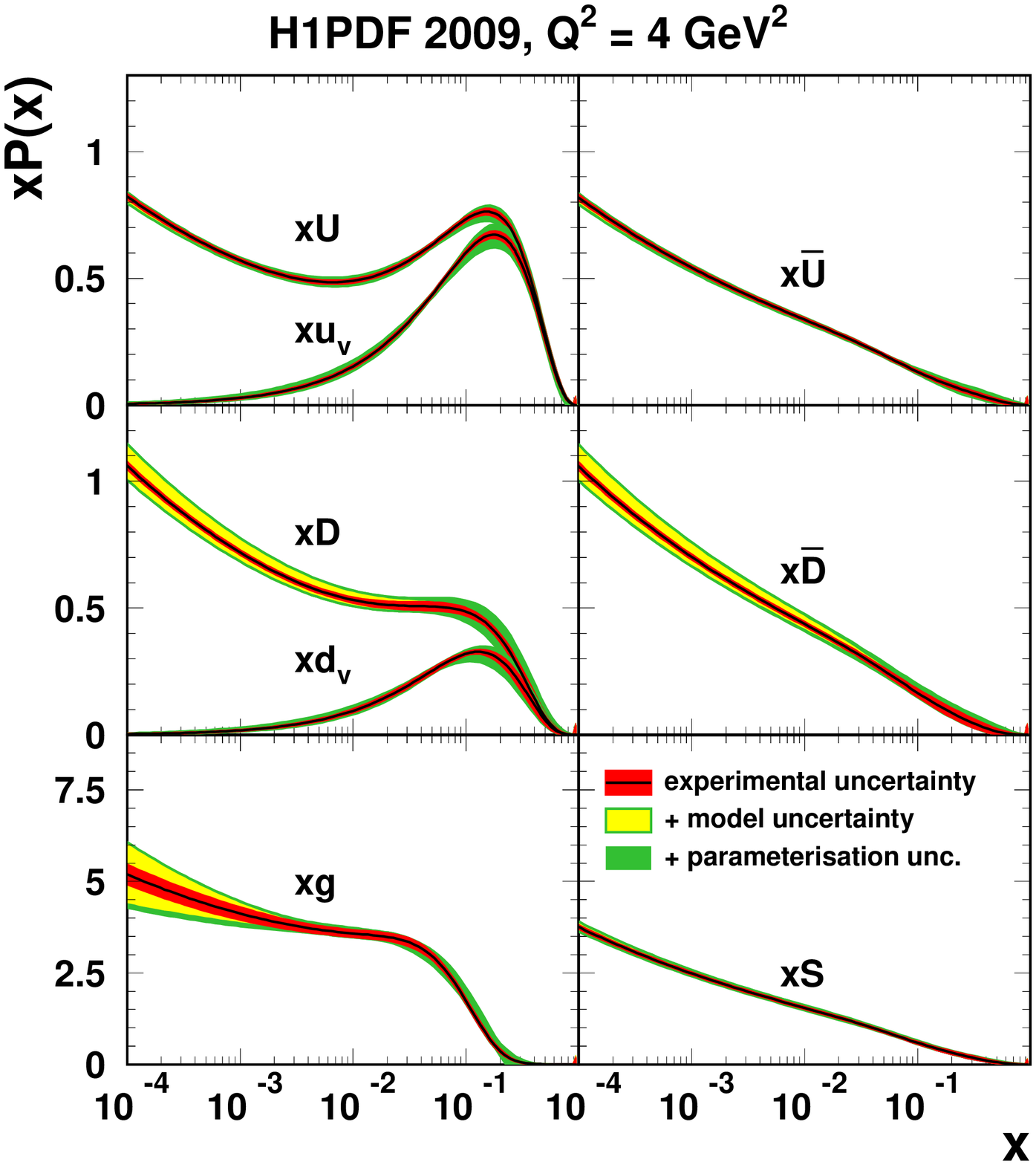}
      \caption{The PDFs from the H1 PDF  fit at $Q^2=4\,\mathrm{GeV}$. }
   \end{center}
    \label{ferrando:fig:qcdfit2}

\end{figure}

\begin{figure}[bt]

  \begin{center}
     \includegraphics[width=.47\textwidth]{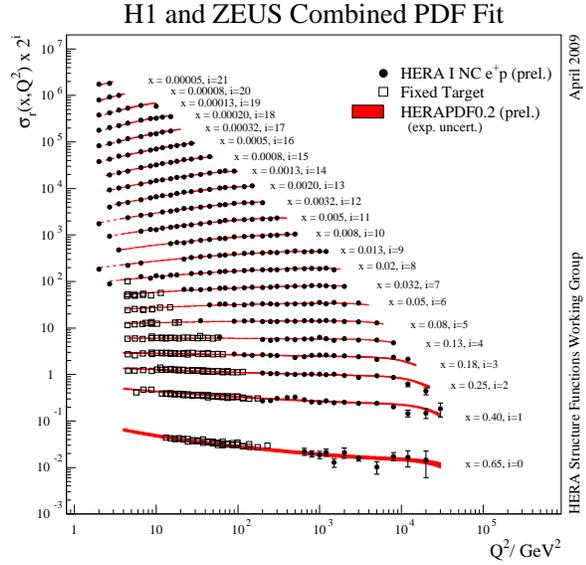}

   \end{center}
      \caption{Results from the H1+ZEUS combined PDF fit compared to HERA-I NC data. }
    \label{ferrando:fig:qcdfit1}

\end{figure}

Both H1~\cite{ferrando:h1:pdf2009} and ZEUS~\cite{ferrando:zeusjets} have performed NLO QCD fits to determine the PDFs of the proton using their data. The latest H1 result is typical of the approach used.   Parametrisations of the following components of the proton are fitted : the up and down valence quarks $xu_v$ and $xd_v$; 
the gluon $xg$, the up-type sea contribution $x\bar{U}=\bar{u}+\bar{c}$ and 
the down-type sea contribution  \mbox{$x\bar{D}=\bar{d}+\bar{s}$}. The parametrisations are determined at a chosen value of $Q^2$, $Q^2_0=1.9\,\mathrm{GeV}^2$.
The parametrised PDFs can then be evolved to different $Q^2$ using the DGLAP equations~\cite{ferrando:dglap}. In these fits, heavy-flavour production is treated according to the general mass variable flavour numbering scheme of Thorne and Roberts~\cite{ferrando:trgmvfns}. The fit uses only H1 data from NC and CC scattering in the HERA-I running period. A good fit to the data is obtained with $\chi^2/\mathrm{n.d.f}=587/644$.  

The PDFs from this fit are shown in Fig. \ref{ferrando:fig:qcdfit2}
The PDF uncertainties are reduced at low-$x$,
compared to the previous H1 fit~\cite{ferrando:h1:pdf2000} (H1PDF2000). 
The 
uncertainties at higher-$x$ are, however, larger than H1PDF2000. These larger
uncertainties are more realistic:  the increase comes from a new parametrisation uncertainty. The parametrisation uncertainty is determined by using 
parametrisations which describe the data well but have unphysical behaviour at
high $x$.

A similar fitting procedure has been adopted to produce a fit to the combined H1+ZEUS data~\cite{ferrando:h1zeus:nccc}. This new fit (HERAPDF0.2) is compared to data from NC scattering in Fig. \ref{ferrando:fig:qcdfit1}. It can be seen that the HERAPDF0.2 gives a very good fit to the data, the $\chi^2/\mathrm{n.d.f}$ for the fit is 
576/592. The precision of the new HERAPDF0.2 is comparable to the global fits
from the CTEQ~\cite{ferrando:cteq66} and MSTW~\cite{ferrando:mstw08} groups. The fit gives much more precisely determined PDFs using the combined data than using the same H1 and ZEUS data separately.

The QCD fits discussed so far were performed using only HERA-I data sets. The
HERA-II data sets include an order of 
magnitude more $e^-p$ data. This
 of should enable a better determination of $xF_3$ and, by extension, the valence
quark distributions. In addition the large (factor 3-4) increase in total
statistics offers improvement in regions where measurements were statistically limited. Finally the new data for the determination of $F_L$ will yield complementary
information about the gluon distribution. These expectations have been verified
by a new QCD fit~\cite{ferrando:zeus:09fit} (ZEUS09) which includes the data used for ZEUS-JETS and the ZEUS HERA-II data described in sections \ref{ferrando:sec:nc} and \ref{ferrando:sec:cc}.

\subsection{Heavy Flavour Production}

\begin{figure}[tb]

  \begin{center}
     \includegraphics[width=.40\textwidth]{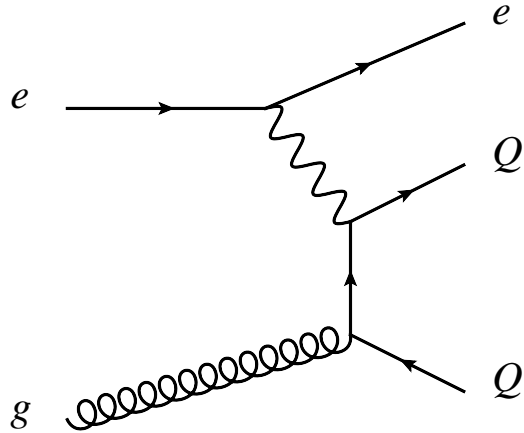}

   \end{center}
      \caption{Production of a heavy flavour quark antiquark pair in DIS via Boson-Gluon Fusion. }
    \label{ferrando:fig:bgf}

\end{figure}

Measurements of heavy-flavour production at HERA have not yet been included
in the QCD fits performed by H1 and ZEUS. Addition of this data could add
further sensitivity to the gluon distribution. This arises because the
main production mechanism for heavy flavour production at HERA is 
boson-gluon fusion (BGF) where a $\gamma$ and gluon produce
a $Q\bar{Q}$ pair (where $Q$ is a heavy quark) as show in Fig. \ref{ferrando:fig:bgf} At HERA, these heavy flavour events can be tagged in three different ways:

\begin{itemize}
\item using heavy flavour-mesons such as the $D^*$;
\item using leptons produced in semi-leptonic decays of the heavy quark;
\item using inclusive impact parameter information. 
\end{itemize}

These different tagging methods provide complementary measurements of heavy flavour production in DIS. They can be directly compared by extrapolating them to the charm  structure function $F_2^{c\bar{c}}$ or beauty structure function $F_2^{b\bar{b}}$.
 In this section two new measurements of $F_2^{c\bar{c}}$ and $F_2^{b\bar{b}}$ are discussed:
one from ZEUS\cite{ferrando:zeus:f2cmu} using semileptonic decays and one from H1\cite{ferrando:h1:f2cip} using inclusive impact parameter distributions.

The new ZEUS result includes the first measurement of $F_2^{c\bar{c}}$ using
semileptonic decays into muons at HERA. A sample of DIS events with a muon and
associated jet, in the kinematic region $Q^2 > 20 \,\mathrm{GeV}^2$, was used. The fractions of such events
arising from charm and beauty events were simultaneously extracted from a fit
to three distributions that distinguish charm and beauty events from each other and  from light quark events.
These variables were: the muon momentum transverse to the axis of the associated jet, the impact parameter of the muon in the transverse plane and the missing transverse momentum parallel to the muon direction. The results were found to agree well with the NLO QCD predictions~\cite{ferrando:hvqdis} and also well with previous measurements that were based on independent techniques.

The new H1 result makes use of inclusive impact parameter distributions to simultaneously fit the charm and beauty contributions to the DIS scattering cross section. The $Q^2$ range $5<Q^2<2000\,\mathrm{GeV}^2$   is  covered.  Although the background from light flavour events is larger than in other methods, a larger region of the phase space for heavy quark induced events is used with this method. This means that the extrapolation from the measured cross sections to $F_2^{Q\bar{Q}}$ is smaller. The measurements agree well with the NLO QCD predictions.

\begin{figure}[t]
\vspace{1.4cm}

   \begin{center}
 \hspace{-2.0cm}\includegraphics[width=.40\textwidth]{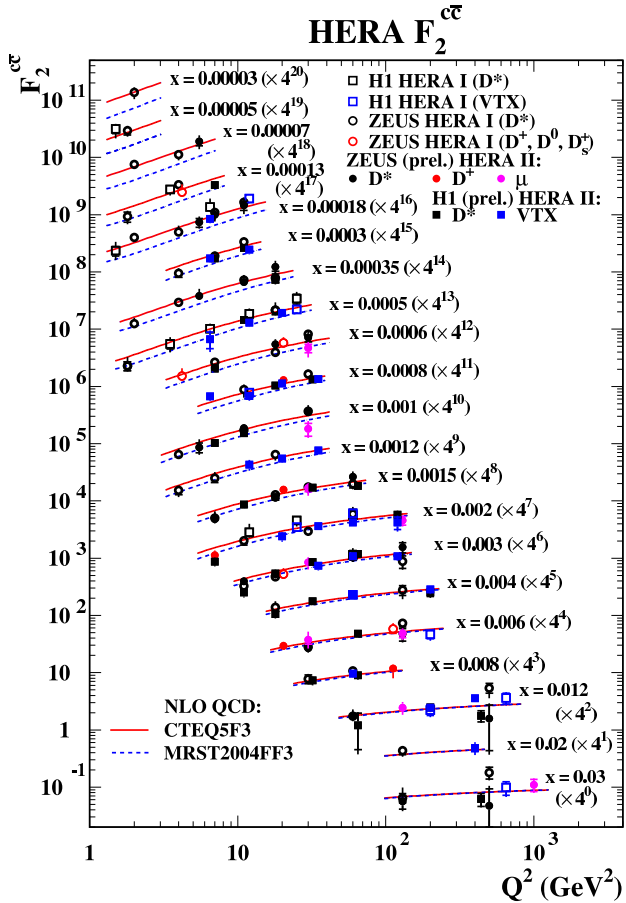}
   \end{center}
      \caption{HERA measurements of $F_2^{c\bar{c}}$. }
    \label{ferrando:fig:hfl}
\end{figure}

A summary of $F_2^{c\bar{c}}$ measurements from H1 and ZEUS is shown in Fig. \ref{ferrando:fig:hfl} Measurements using all three heavy flavour tagging methods are shown. It can be seen that a detailed and consistent picture of $F_2^{c\bar{c}}$ has
been revealed by HERA. There is much smaller coverage of $F_2^{b\bar{b}}$, but more measurements are still expected from both H1 and from ZEUS. Inclusion of these measurements in QCD fits of the PDFs would provide, at the very least, an important consistency check for the gluon distribution. Combination of the H1 and ZEUS measurements is currently underway, and strong improvements in precision are expected.

\subsection{Isolated $\gamma$ Production}

\begin{figure}[t]

   \begin{center}
     \includegraphics[width=.49\textwidth]{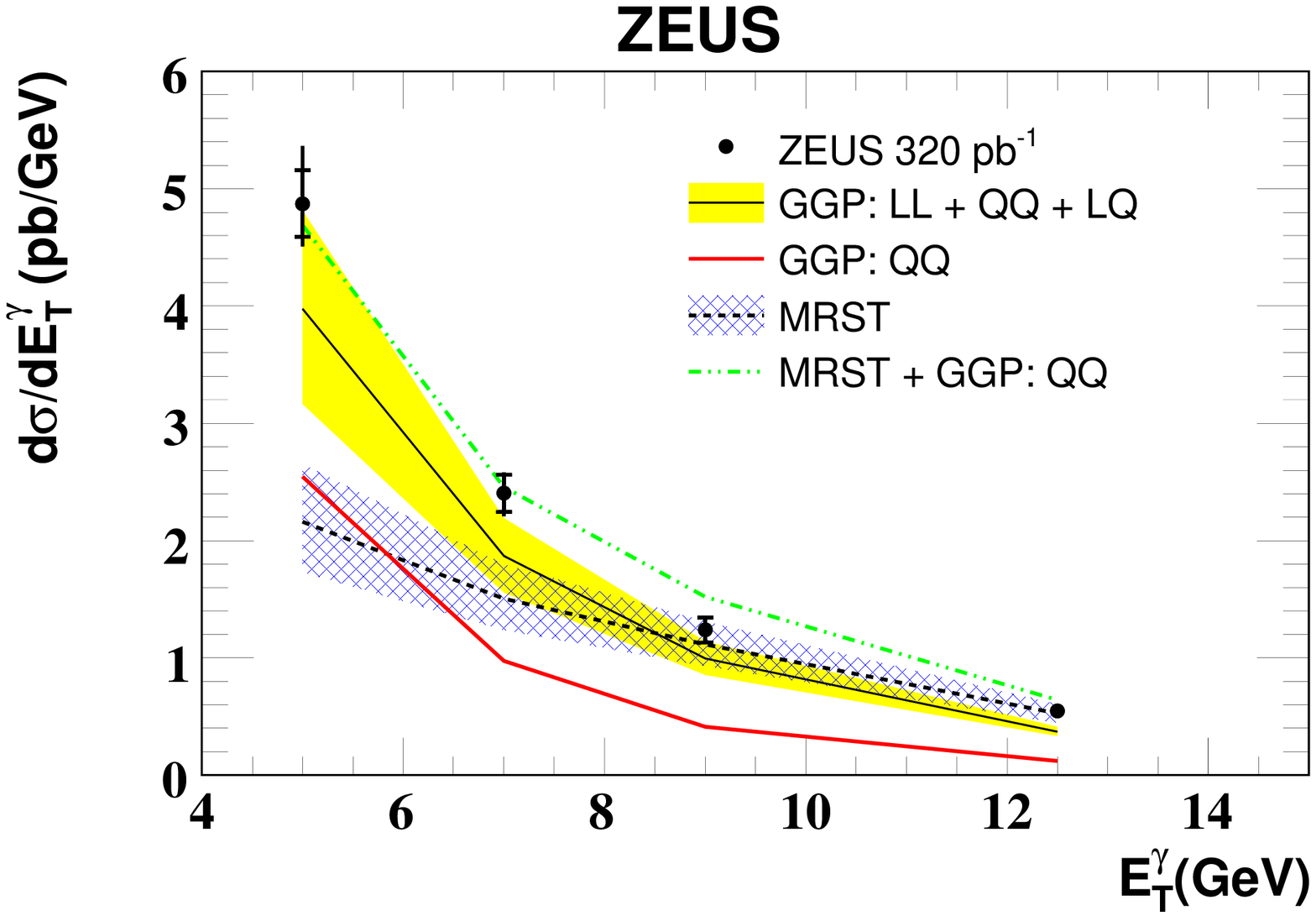}
     \includegraphics[width=.49\textwidth]{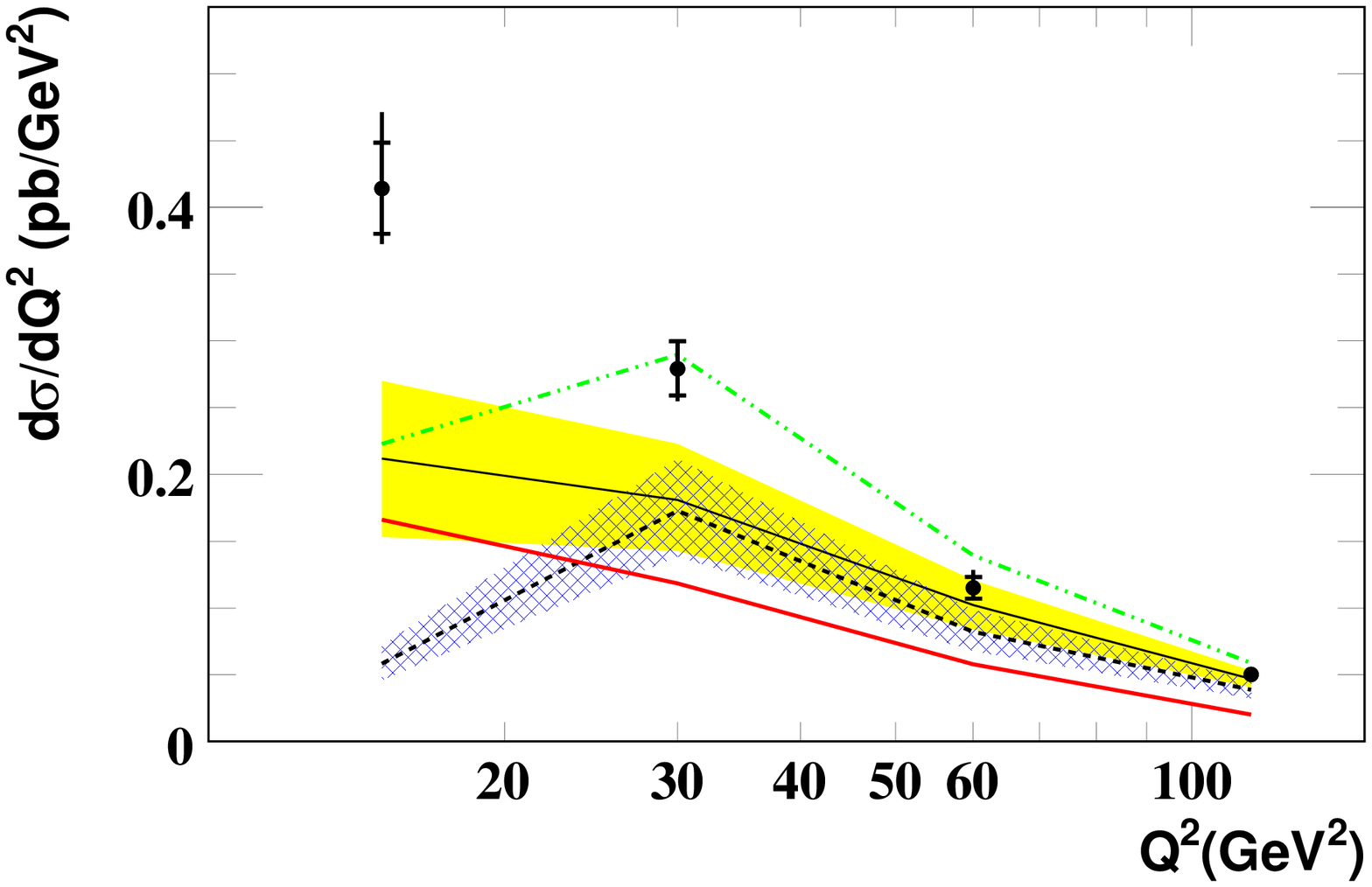}
   \end{center}
      \caption{Differential cross sections for isolated photon production in DIS from the ZEUS collaboration. }
    \label{ferrando:fig:isophot}
\end{figure}

Isolated photons can be produced in DIS via photon radiation from the lepton 
or from a quark. The contributions to the cross section from lepton radiation 
(LL), quark radiation  (QQ) and  the interference (LQ) between them have
 been calculated at leading order in Quantum electrodynamics (QED) and QCD by 
Gehrmann-De Ridder {\em et al.}~\cite{ferrando:ggp} (GGP). However, Martin {\em et al.} (MRST) have 
shown that an enhancement of the LL contribution would arise as a consequence
 of adding QED corrections to PDFs~\cite{ferrando:mrst04}. In this case the proton gains a photonic component ($\gamma_P$) which then undergoes Compton scattering $e\gamma_P \rightarrow e\gamma$\footnote{This photonic component of the proton would have consequences for calculations of electroweak corrections to cross sections at the LHC and TeVatron.}. The measurement of this cross section is hence sensitive to the photonic component of the proton.

A new measurement of isolated photon production in DIS from the ZEUS experiment~\cite{ferrando:zeus:isophot} in the kinematic range $10 < Q^2 < 350\,\mathrm{GeV}$ has
been made. The measured differential cross sections as functions of the photon
transverse momentum, $E_T^{\gamma}$ ($d\sigma/dE_T^{\gamma}$) and $Q^2$ ($d\sigma/dQ^2$) are compared to the theoretical calculations in Fig.~\ref{ferrando:fig:isophot} Since the MRST calculation does not include the QQ part of the cross section an hybrid calculation can be made by summing it with the QQ part of the GGP calculation (the LQ term can be neglected).
It can be seen that the shape of the $d\sigma /dE_T^{\gamma}$ is well described both by GGP and the hybrid calculation (labelled MRST + GGP:QQ). However the GGP calculation significantly underestimates
the differential cross section at low $Q^2$, as was also observed in a recent measurement by H1~\cite{ferrando:h1:isophot}. The hybrid calculation improves the description of $d\sigma/dQ^2$, but still underestimates the differential cross section at lowest $Q^2$. A version of the MSTW08 PDFs~\cite{ferrando:mstw08} incorporating QED corrections is under preparation and may offer an improvement in description of the data.

\section{Jets}

Jet measurements at HERA are a precision test of QCD. The measured cross sections depend on the gluon content of the proton and on the magnitude of the strong coupling. Statistical precision is typically very good and, experimentally, 
the usual limitation on precision comes from  the jet-energy-scale uncertainty. In this section recent results from jet measurements in high-$Q^2$ DIS are
reviewed.

\subsection{Jet production in DIS}
\label{ferrando:sec:jetmeas}

\begin{figure}[t]

   \begin{center}
     \includegraphics[width=.50\textwidth]{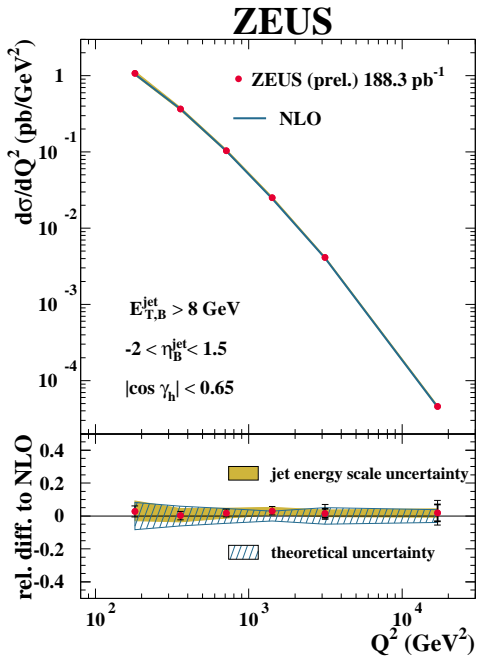}
   \end{center}
      \caption{The differential cross section for inclusive jet production in high-$Q^2$ DIS, measured by the ZEUS collaboration. }
    \label{ferrando:fig:zeusncjet}
\end{figure}

The ZEUS collaboration have recently measured the cross sections for inclusive
jet production in DIS at high-$Q^2$ ($Q^2 > 125\,\mathrm{GeV}^2$)~\cite{ferrando:zeus:hq2jets}. The analysis uses a sample of NC DIS events containing at least one jet with $E_T$ in the Breit frame~\cite{ferrando:dcs} greater than $8\,\mathrm{GeV}$ and
pseudorapidity in the Breit frame within the region $-2 < \eta^{\mathrm{jet}} < 1.5$. The measured differential cross section as a function of $Q^2$ is compared to NLO QCD predictions from the programme DISENT~\cite{ferrando:theory:jet} in Fig.~\ref{ferrando:fig:zeusncjet} 
It can be clearly seen that the measurements agree well with the predictions.
It can also be seen that the precision of the measurement is often better than
that of the predictions. In the regions where this is true, the uncertainty on the prediction is often dominated by the uncertainty on the gluon PDF, thus
this measurement can constrain the gluon PDF if used as an input to a PDF fit.

\begin{figure}[t]


   \begin{center}
     \includegraphics[width=.47\textwidth]{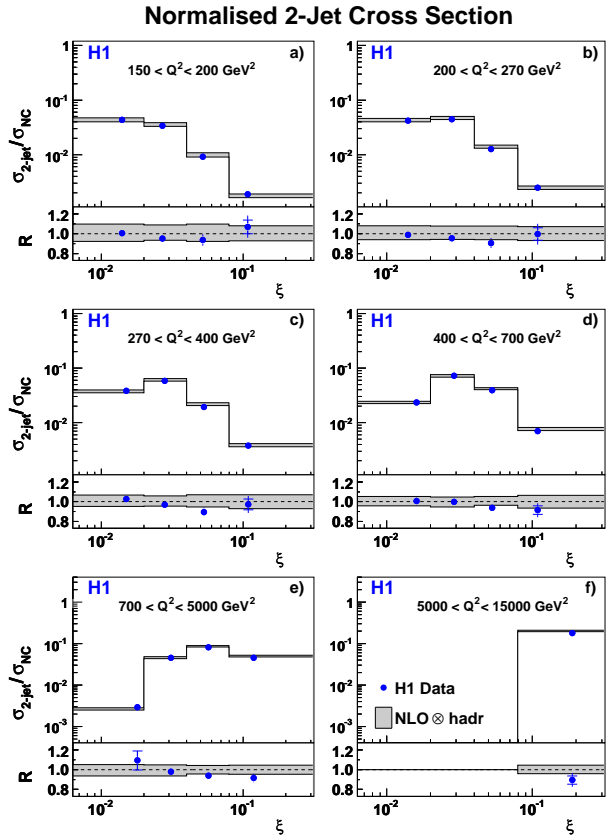}
   \end{center}
      \caption{The normalised cross section for dijet production in High-$Q^2$ DIS, measured by the H1 collaboration. }
    \label{ferrando:fig:h1ncjet}
\end{figure}

The H1 collaboration has also recently presented a new measurement of 
inclusive jet production in high-$Q^2$ DIS ($150< Q^2 < 15000\,\mathrm{GeV}^2$). This measurement is presented alongside a measurement of the normalised cross sections (e.g. the ratios to the fully inclusive NC cross section)  for two- and three-jet production. The normalised dijet cross section is compared to predictions from DISENT and NLOJET++~\cite{ferrando:nlojet} in Fig. \ref{ferrando:fig:h1ncjet} The comparison is shown as a function of $\xi$, the momentum fraction carried by the incoming parton: $$ \xi= x(1+\frac{M_{12}}{Q^2}), $$ where $M_{12}$ is the dijet invariant mass. It is clear that the H1 measurements are significantly more precise than the  predictions over a large part of the kinematic range. In this case the dominant uncertainties on the predictions come from missing higher order terms for the calculation.

\subsection{Measurements of the strong coupling}

\begin{figure}[t]

   \begin{center}
     \includegraphics[angle=270,width=.45\textwidth]{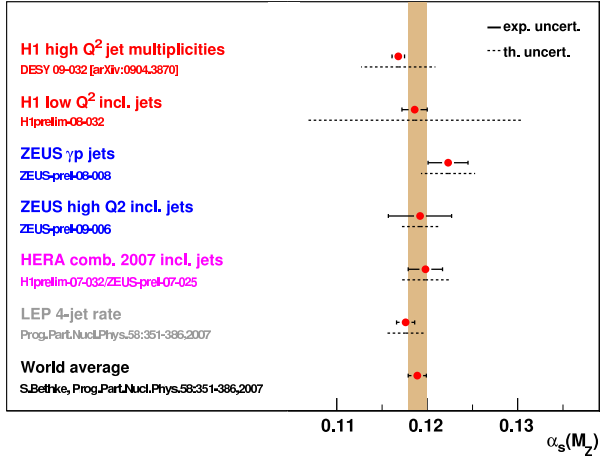}
   \end{center}
      \caption{A summary of $\alpha_S$ measurements from HERA.
    \label{ferrando:fig:heraalphas}}
\end{figure}

Jet cross-section measurements at HERA are sufficiently precise to enable
measurement of the strong coupling $\alpha_S$. This is done by performing
the theoretical calculation using different values of $\alpha_S$ (and appropriate PDF sets) in order to make a phenomenological parametrisation of
the cross section as a function of $\alpha_S$. This establishes a direct
mapping from a measured cross section  to $\alpha_S$ with
associated uncertainty. A summary plot of selected measurements of $\alpha_S$, including
those described in section~\ref{ferrando:sec:jetmeas}, is shown in Fig \ref{ferrando:fig:heraalphas} The H1 measurement at high $Q^2$ is of comparable experimental precision to the measurement  from four-jet events at LEP and to the world average~\cite{ferrando:qcd:wav}. The impressive success of QCD is visible, with consistent measurements from differing environments and energy regimes.

\subsection{Jet Substructure}

The study of the substructure of jets can provide information about the pattern of radiation from the primary parton initiating the jet. In $ep$ collisions, substructure can also be used to explore colour coherence  by studying to what extent the soft radiation is emitted in the direction of the proton.
Jet substructure is usually studied by defining jets using a certain algorithm
e.g the $k_T$ algorithm~\cite{ferrando:ktalg}, and then running a different algorithm
(or the same algorithm with different parameters)  on the constituents of the jets found by the original algorithm. Recently jet substructure has been suggested as a useful tool for searching for boosted heavy particles at the LHC~\cite{ferrando:lhcsubs}.
Any such usage will be contingent on the substructure of jets from background QCD processes being well understood.

\begin{figure}[t]

   \begin{center}
     \includegraphics[width=.49\textwidth]{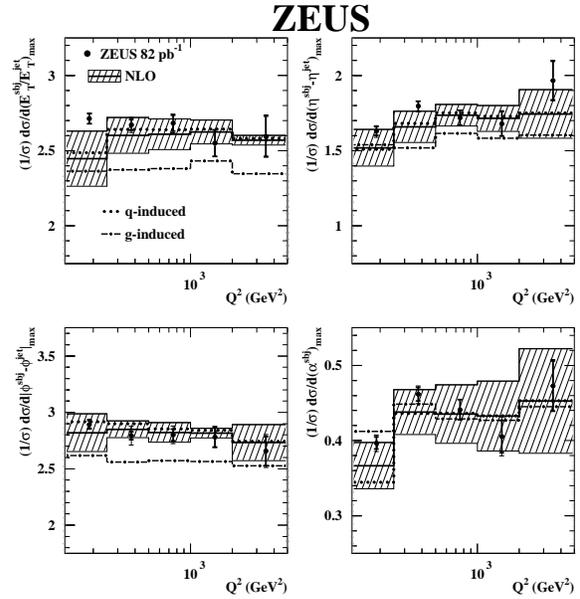}
   \end{center}
      \caption{Maxima of the normalised two-subjet differential cross sections as a function of $Q^2$ compared to QCD calculations. for comparison, NLO predictions for gluon and quark indiced jets are also shown separately.
    \label{ferrando:fig:zeussub}}

\end{figure}

The ZEUS collaboration has recently produced two new substructure measurements, one of jets with exactly two subjets~\cite{ferrando:zeus:subs1} and one of jets with exactly three subjets~\cite{ferrando:zeus:subs2}. In both cases jets are first selected by running the $k_T$ algorithm in the longitudinally invariant inclusive mode. The subjet analysis is made by running the exclusive $k_T$ algorithm on 
the constituents of these jets using a resolution parameter $y_{\mathrm{cut}}$ of 0.05 (the two subjets measurement) or 0.03 (the three subjets measurement). The measurement kinematic
region $Q^2 > 125 \,\mathrm{GeV}^2$ was used for both measurements. The measured normalised subjet cross sections for the two-subjet analysis as a function of $Q^2$ are compared to predictions from DISENT in Fig \ref{ferrando:fig:zeussub}. The results agree adequately with the perturbative QCD predictions over a large range of $Q^2$. the precision of the data is often better than that of the predictions. The measurements of jets with three-subjets leads to similar conclusions. The
capability of pQCD to describe QCD jet substructure is encouraging for application of such infromation in new physics searches.

\section{Summary}
With data taking complete, the HERA experiments continue to produce important QCD measurements. HERA produces world leading PDF measurements: the newest data at high $Q^2$ will help to further constrain quark PDFs, while jet and heavy flavour messages yield important information about the gluon density. Isolated photon events in DIS also offer the opportunity to probe the photonic component of the proton. Jet production at HERA is a rich testing ground for QCD, 
extremely precise measurements of the strong coupling have been made and
the precision of jet cross sections is often limited by theoretical input.
As attention at the LHC turns to jet substructure as a tool for new physics
searches, measurements of these quantitities at HERA can help us to assess
our understanding of this aspect of QCD. Not content with these separate precise QCD measurements, the H1 and ZEUS collaborations have worked together to produce a combination technique for their data that provide structure function measurements of unsurpassed precision.

\end{document}